# Singular teleparallelism

D. H. Delphenich
Spring Valley, OH 45370
e-mail: feedback@neo-classical-physics.info

**Abstract**: It is shown that the geometry of parallelizable manifolds can be extended to non-parallelizable ones by extending the connection that a global frame field would define on a parallelizable manifold to a connection that a singular frame field would define on a non-parallelizable one. The resulting connection would typically have non-vanishing curvature in the neighborhood of the singular points of the frame field. The example of a 2-sphere is discussed as a motivating example and later extended to more general suspensions of parallelizable manifolds.

**1. Introduction.** – Once Einstein's theory of gravitation (viz., general relativity) started gaining some acceptance in the early Twentieth Century, the fact that it was so deeply rooted in the non-Euclidian geometry of the space-time manifold made many theoretical physicists, including Einstein himself, wonder if Riemannian geometry was as far as things went or was there a non-Riemannian geometry that would include the geometry of relativity, and perhaps explain other physical phenomena besides gravitation. In particular, Einstein and others wondered if it might be possible to also find a geometric explanation for electromagnetism. Since the Levi-Civita connection of Riemannian geometry is defined to be a metric connection with vanishing torsion, the obvious extensions were to non-metric connections, such as conformal connections, which was mostly due to Weyl [**1**], and to non-vanishing torsion, which was developed considerably by Cartan [**2**].

One of Einstein's earliest attempts at unifying his theory of gravitation with Maxwell's theory of electromagnetism took the form of what came to be called "teleparallelism," or "distant parallelism" or "absolute parallelism" or even the geometry of "Weitzenböck spaces." Whereas the Riemannian approach to parallel translation is local, and typically defined only along curves, in teleparallelism, one starts with the existence of a global frame field on one's manifold and uses it to define parallelism of geometric objects (i.e., tensors) at finitely-separated points by demanding that they must have the same components with respect to the frames that are defined at those points.

The field theory that Einstein and others ([1]) derived by using the global frame field as the fundamental field, rather than the space-time metric, eventually proved to be unsatisfactory, since (among other things) it admitted unphysical solutions, such as a static configuration of gravitating bodies.

It is intriguing that the point in time (viz., the early 1930's) that interest in teleparallelism started declining was still a few years prior to the first definitive work in which the topological obstructions to global parallelizability were first addressed. In 1935, Hassler Whitney [**4**] derived some $\mathbb{Z}_2$-cohomology classes that related to parallelizability by looking at sphere bundles (such as all unit vectors in the tangent bundle to a Riemannian manifold). In the next year, Eduard Stiefel published a doctoral dissertation under Heinz Hopf [**5**] that was entitled (in translation) "Direction fields and teleparallelism in *n*-dimensional manifolds" and expanded upon the problem of topological obstructions to parallelizability in more detail than the previous brief note by Whitney. Stiefel's thesis was explicitly an attempt to extend the Poincaré-Hopf theorem [**6**], which related the existence of non-zero vector fields on a manifold to the vanishing of its Euler-Poincaré characteristic, to the existence of more than one linearly-independent non-zero vector field on the manifold, and to characterize the maximum number of such vector fields that one could have. In that thesis, Stiefel defined some $\mathbb{Z}_2$-*homology* classes whose vanishing would be necessary (but not sufficient) for the parallelizability of a differentiable manifold. The two approaches to the derivation of obstructions that were taken by Whitney and Stiefel were seen to be equivalent, and today one typically refers to the *Stiefel-Whitney* classes of the manifold (or really its bundle of linear frames). (For a more modern discussion of those classes, see, e.g., Milnor and Stasheff [**7**].)

---

([1]) The author has compiled a collection of his English translations of many of the early papers on teleparallelism by Einstein, Mayer, Cartan, Bortolotti, and Zaycoff, among others, in a book [**3**] that is available as a free PDF download at his website: neo-classical-physics.info.

Some of the reason for the topological oversight in early teleparallelism was simply the fact that differential geometry in the time when Einstein was developing general relativity was pure local in character. The modern topological definition of a differentiable manifold simply did not exist yet, so it was not commonplace for people to consider the degree to which topology might obstruct geometry. Indeed, Haussdorff's definition of a topological space was being introduced at roughly the same time as general relativity.

As a result of that historical state of affairs, this author has always been curious about how the theory of teleparallelism might have changed had Einstein, Cartan, and others been more consciously aware of the topological ramifications of parallelizability. In particular, is there some way of extending the geometry (viz., the connection) of teleparallelism to non-parallelizable manifolds, even though one clearly cannot extend the frame field? Perhaps the topological information in the neighborhood of singular points where the frame field cannot be extended might lead to non-vanishing curvature, whereas the connection of any non-singular frame field will always have vanishing curvature.

This paper is intended to address precisely that issue. It represents essentially a continuation of a previous paper [**8**] on the purely geometric aspects of teleparallelism, the contents of which will be assumed in what follows. Although some comments on the nature of singular teleparallelism were made in the author's introduction to [**3**], nonetheless, some details had yet to be developed at that point in time, and this paper will serve as a first attempt to define the extended teleparallelism connection when the fundamental frame field has singular points.

Section **2** begins by defining the concept of a singular frame field and citing various examples of parallelizable manifolds, along with quoting a useful theorem for constructing manifolds of a given degree of parallelizability. Section **3** discusses the motivating example for the following basic construction, namely, how to extend the teleparallelism connection on a 2-sphere minus the poles. Section **4** then develops the core of the present study, namely, a particular technique for extending a more general teleparallelism connection in the neighborhood of singular points and the geometric consequences of that definition. In Section **5**, a broad class of (typically) non-parallelizable manifolds that includes the 2-sphere and 4-sphere is defined by the suspensions of parallelizable manifolds and they are shown to admit the aforementioned extensions. Finally, some issues that follow from the present study are discussed in the last section.

Although the subject of this article is basically topological in character, nonetheless, the level of topology that will be employed is mostly point-set theoretic in character. Hence, one might learn most of it from the text by Munkres [**9**].

**2. Singular frame fields.** – There are essentially two different ways of introducing singular points into frame fields: introducing zeroes into the individual vector fields that comprise the frame members and introducing points at which the frame fields are not defined. One can use the 2-sphere as an illustrative example for both approaches.

The Poincaré-Hopf theorem [**6**] says that a compact manifold $M$ will admit a non-zero vector field iff its Euler-Poincaré characteristic $\chi[M]$ vanishes. For instance, that will be true of any odd-dimensional sphere, so any vector field on the 2-sphere must have at least one zero. For the purposes of computation and illustration, it is generally more convenient to let the vector field have two zeroes, and one typically locates them at the North and South poles.

In general, the set of points $Z(\mathbf{v})$ in $M$ at which a vector field $\mathbf{v}(x)$ has a zero does not have to tell one anything about the topology of $M$. That is because one can always extend (possibly with the use of a "bump function") a non-zero vector field that is defined on a proper subset of $M$ to a global vector field by way of zeroes, which will not be affected by the topology of the manifold outside of the original support for the vector field. Hence, one must define a class of vector fields with zeroes that are "generic" to the topology of $M$ in the sense that $Z(\mathbf{v})$ is indeed determined by the topology of $M$. Our definition of genericity will be that $\mathbf{v}$ is *generic* iff $Z(\mathbf{v})$ is minimal in the partial ordering of subset inclusion as $\mathbf{v}$ ranges over all vector fields on $M$. Hence, there will be no vector field $\mathbf{w}$ for which $Z(\mathbf{w})$ is a proper subset of $Z(\mathbf{v})$. A particular tractable example is when $Z(\mathbf{v})$ consists of a finite set of isolated points; in some cases, it could be just one point, such as with the 2-sphere. Note that the set of zeroes of any (continuous) vector field $\mathbf{v}(x)$ on a differentiable manifold $M$ will be a closed subset $Z(\mathbf{v})$ of $M$, regardless of its genericity.

A set of $k$ vector fields $\{\mathbf{e}_1(x), \ldots, \mathbf{e}_k(x)\}$ on an $n$-dimensional differentiable manifold $M$ such at any point $x \in M$, either at least one of them has a zero or they are all linearly-independent will be called a *singular k-frame field* on $M$. For instance, when one puts two vector fields $\{\mathbf{e}_1(x), \mathbf{e}_2(x)\}$ on the 2-sphere that each have zeroes at their singular points, but are otherwise linearly-independent at each non-singular point, one can regard the set $\{\mathbf{e}_1(x), \mathbf{e}_2(x)\}$ as a singular 2-frame field on the 2-sphere.

If one has a singular $k$-frame field $\{\mathbf{e}_1, \ldots, \mathbf{e}_k\}$ then the set $Z(\mathbf{e}) = Z(\mathbf{e}_i) \cup \ldots \cup Z(\mathbf{e}_i)$ of all singular points of the individual vector fields will be the finite union of closed sets and therefore closed. The question of its minimality relates to the way that the sets $Z(\mathbf{e}_i)$ intersect and whether one could consolidate singularities at the same points. For instance, in the case of the 2-sphere, one can arrange that the two vectors fields that define a singular frame field have their zeroes at the same point. If all of the vectors of a singular frame field are generic with a finite number of isolated singular points then they will have either no singular points or the same number of singular points. One could typically arrange that those singular points should overlap.

Once again, we will call the singular *k*-frame field *generic* iff $Z(\mathbf{e})$ is minimal. From now on, vector fields and frame fields will typically be assumed to be generic.

If the manifold *M* has a positive-definite metric *g* on its tangent bundle, so that one can speak of normalizing vector fields, then when one divides one of the frame members $\mathbf{e}_i(x)$ by its norm $\|\mathbf{e}_i(x)\|$ at each non-singular point *x*, the resulting vector field will be a unit vector field everywhere except the zeroes where it will be undefined. More to the point, one can orthonormalize the frame members at every non-singular point, which will then make:

$$g(\mathbf{e}_i(x), \mathbf{e}_j(x)) = \delta_{ij} \tag{2.1}$$

at any non-singular point *x*.

The two approaches to defining a singular point are illustrated in Figs. 1.*a* and 1.*b*.

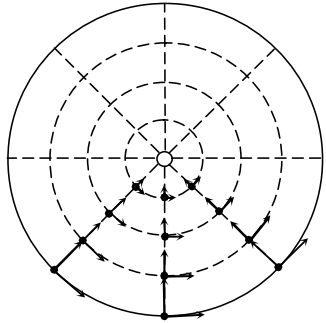 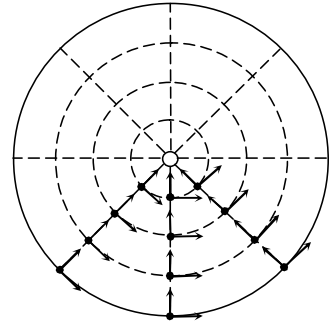

Figure 1.*a*. – Singularities as zeros.    Figure 1.*b*. – Singularities as undefined points.

Interestingly, the approach that Stiefel took to the description of the singularities of frame fields on general (i.e., not-necessarily-parallelizable) manifolds in his thesis under Hopf [4] was to assume that the singularity subset $Z(\mathbf{e})$ of a *k*-frame field $\{\mathbf{e}_1, \ldots, \mathbf{e}_k\}$ could be triangulated into a simplicial complex. It could then be expressed as a sum of simplexes of various dimensions with integer coefficients. He then showed that if one reduced the coefficient ring from $\mathbb{Z}$ to the ring $\mathbb{Z}_2 = \{0, 1\}$, which means that if a coefficient is odd then it will get replaced with 1 and if it is even then it will get replaced with 0 (i.e., dropped from the sum), then the formal sum of all the simplexes in the complex with $\mathbb{Z}_2$ coefficient would be closed in the sense of having a vanishing boundary, as would each term in the sum. One then gets a series of $\mathbb{Z}_2$-homology classes that are now referred to as the *Stiefel-Whitney* classes (or rather their dual $\mathbb{Z}_2$-cohomology classes). Just as the zeroes of vector fields on a compact manifold answer to the Euler-Poincaré characteristic of the (tangent bundle to that) manifold (viz., the alternating sum of the Betti numbers), the singularities of frame fields must answer to those Stiefel-Whitney classes.

In particular, one can speak of the *degree of parallelizability* of a manifold, which is the maximum number of non-zero vector fields that one can define on it that will still be linearly-independent at all points. One already sees that the definition implicitly assumes that non-zero vector fields even exist on the manifold to begin with. Hence, the 2-sphere could be said to have a degree of parallelizability zero, while a parallelizable manifold of dimension $n$ would have the maximum degree of parallelizability, namely, $n$. On the other hand, a sphere with a vanishing Euler-Poincaré characteristic (such as odd-dimensional ones) will admit a non-zero vector field, so it will have a degree of parallelizability that is equal to at least 1.

Examples of parallelizable manifolds are, on the one hand, restricted in scope, but on the other hand, they include many elementary examples that are still quite useful to consider. For instance, affine spaces, as well as vector spaces, are parallelizable, as are all Lie groups. According to Stiefel [**5**], all compact, orientable 3-manifolds are parallelizable, such as both the 3-sphere and three-dimensional projective space. However, most of the spheres are non-parallelizable, except for the ones of dimension 0, 1, 3, and 7. As it happens, the spheres of dimension 0, 1, and 3 are also the underlying manifolds for Lie groups, namely, $\mathbb{Z}_2$, $SO(2)$ or $U(1)$, and $SU(2)$, resp.

Some simple examples of the intermediate cases of the degree of parallelizability can be obtained from the:

**Theorem:**

*The product $M \times N$ of a manifold $M$ of degree of parallelizability $p$ with one $N$ of degree of parallelizability $q$ will have degree of parallelizability $p + q$.*

**Proof:**

If $M$ has degree $p$ then a $p$-frame field $\{\mathbf{e}_1(x), \ldots, \mathbf{e}_p(x)\}$ will exist on it. Similarly, a $q$-frame field $\{\mathbf{f}_1(x), \ldots, \mathbf{f}_q(x)\}$ will exist on $N$. Since the tangent spaces $T_x M$ and $T_y N$ to $M$ and $N$, resp., are complementary subspaces in the tangent spaces to $T_{(x,y)}(M \times N)$, those two sets of vector fields will be linearly-independent. Hence, the union of the two sets of vector fields will define a $p + q$-frame field on $M \times N$.

In particular:

**Corollary:**

*The product of two parallelizable manifolds is parallelizable.*

Any torus is an obvious example of that fact. Of course, any torus also the underlying manifold of a Lie group, so that would also make it parallelizable. Since we just said that any sphere with non-vanishing Euler-Poincaré characteristic (such as even-dimensional ones) will have degree of parallelizability zero, in order to get an example of an intermediate degree of parallelizability between zero and the maximum, one can form products of $S^{2k}$ with parallelizable manifolds whose dimensions are equal to the desired degree of parallelizability.

**3. A motivating example.** – Our motivating example for the more general construction involves perhaps the simplest of non-parallelizable manifolds, namely, the 2-sphere. If $N$ and $S$ are the North and South poles, respectively, then one usually defines a singular right-handed, orthonormal 2-frame field on $S^2 - \{N, S\}$ by way of the unit vector field $\mathbf{e}_1$ that is tangent to the latitude circles (pointing East) and the one $\mathbf{e}_2$ that is tangent to the longitude circles (pointing North). When one looks down upon the North and South poles, one gets pictures of the kind that are illustrated in Figs. 2.$a$ and 2.$b$:

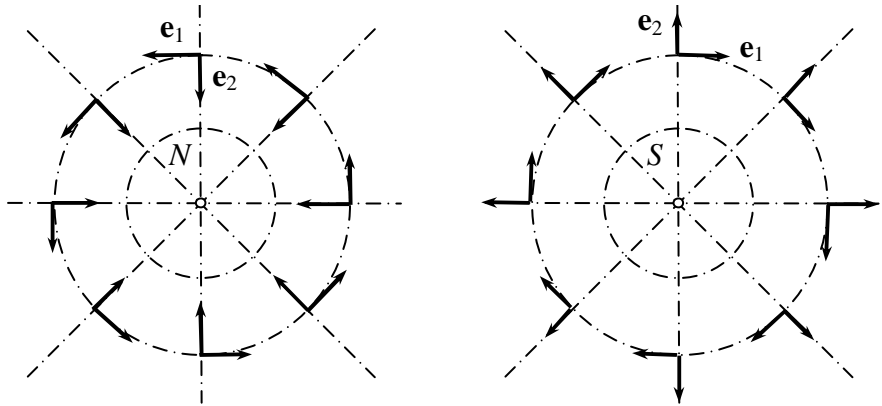

Figure 2.$a$ – View from the North pole.  Figure 2.$b$ – View from the South pole.

If one locates a polar coordinate chart $(r, \theta)$ about each singular point then the resulting frame field that the coordinates define will take the form that is depicted in Fig. 3:

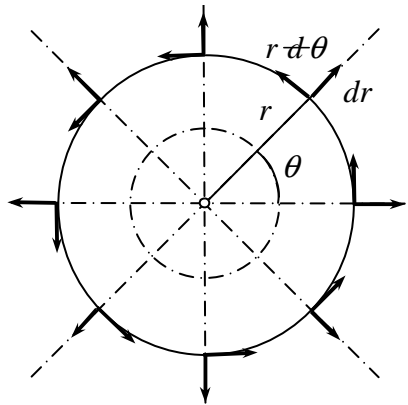

Figure 3. – Polar coordinates.

The reason that the reciprocal unit 1-form that points in the direction of increasing $\theta$ is written $d\!\!\!/\theta$, instead of $d\theta$, as is commonplace is a learned borrowing from thermodynamics, in which the more conscientious authors explicitly distinguish 1-forms that are not exact from ones that are by the use of the slash through the $d$. Indeed, one must note that the angle "function" $\theta$ is not a true function on $\mathbb{R}^2 - \{0\}$, since it is multi-valued along the positive $x$-axis. Hence, it is not rigorous to speak of $d\theta$ as if there were such a function.

However, one does have that $d\theta$ is closed:

$$d_\wedge d\theta = 0. \tag{3.1}$$

The fact that one can have a 1-form on $\mathbb{R}^2 - \{0\}$ that is closed, but not exact, is due to the fact that the space is not simply-connected. Indeed, it has the circle as a deformation retract, which has a fundamental group $\pi_1(S^1)$ that is isomorphic to $\mathbb{Z}$, and thus has one generator. That generator then becomes the basis vector for the one-dimensional vector space $H^1_{dR}(\mathbb{R}^2 - \{0\})$, which is the one-dimensional de Rham cohomology space; i.e., the vector space of all closed 1-forms on $\mathbb{R}^2 - \{0\}$ that are not exact.

The relationship between the polar coframe field $\{dr, r\, d\theta\}$, which we shall call $\{\hat{\theta}^1, \hat{\theta}^2\}$, resp., and the Cartesian one $\{dx, dy\}$, which we shall call $\{dx^1, dx^2\}$, is a simple rotation in the plane through the angle $\theta$:

$$\hat{\theta}^i = R^i_j(x)\, dx^j. \tag{3.2}$$

*a. The basic teleparallelism connection.* – First, let us consider the situation in the neighborhood of $N$ (but not at $N$), where one has:

$$\theta^1 = -\hat{\theta}^1 = -dr, \qquad \theta^2 = \hat{\theta}^2 = r\, d\theta, \tag{3.3}$$

and therefore:

$$\theta^i = \begin{bmatrix} -1 & 0 \\ 0 & 1 \end{bmatrix} R\, dx^i, \quad \text{so} \quad dx^i = \tilde{R}\begin{bmatrix} -1 & 0 \\ 0 & 1 \end{bmatrix} \theta^i. \tag{3.4}$$

Differentiation gives:

$$d\theta^i = \begin{bmatrix} -1 & 0 \\ 0 & 1 \end{bmatrix} dR \otimes dx^i = -\omega^i_j \otimes \theta^j, \tag{3.5}$$

in which

$$\omega^i_j = \begin{bmatrix} 1 & 0 \\ 0 & -1 \end{bmatrix}\begin{bmatrix} 0 & 1 \\ -1 & 0 \end{bmatrix}\begin{bmatrix} -1 & 0 \\ 0 & 1 \end{bmatrix} d\theta = \begin{bmatrix} 0 & 1 \\ -1 & 0 \end{bmatrix} d\theta = \frac{1}{r}\begin{bmatrix} 0 & 1 \\ -1 & 0 \end{bmatrix} \theta^2. \tag{3.6}$$

This is the teleparallelism connection that is defined by the frame field $\mathbf{e}_i$; i.e., the one that makes $\mathbf{e}_i$ parallel. (See [**8**] for a more extended discussion of the basic definitions of teleparallel geometry.) Since the complement of $\{N, S\}$ in $S^2$ is diffeomorphic to a cylinder, which is the product of parallelizable manifolds, the region of $S^2$ on which $\mathbf{e}_i$ is defined is indeed parallelizable.

*b. The structure equations of the connection.* – The torsion of the connection $\omega$ is obtained from direct calculation:

$$\Theta^1 = d_\wedge \theta^1 + \omega^1_i \wedge \theta^i = \omega^1_2 \wedge \theta^2 = 0,$$

$$\Theta^2 = d_\wedge \theta^2 + \omega_i^2 \wedge \theta^i = -\frac{1}{r}\theta^1 \wedge \theta^2 + \frac{1}{r}\theta^1 \wedge \theta^2 = 0 \,;$$

i.e.:
$$\Theta^i = 0. \qquad (3.7)$$

Similarly, the curvature vanishes:

$$\Omega = d_\wedge \omega + \omega \wedge \omega = 0 + 0 = 0 \,, \qquad (3.8)$$

as it should, in general.

For the time being, regard the plane that is described by polar coordinates as the complex line, so that one can write:

$$\omega = i \, d\theta = \frac{i}{r}\theta^2. \qquad (3.9)$$

If $z_1$ is a 1-cycle (i.e., loop) that encircles $N$ (such as the unit circle) then the integral of $\omega$ around that 1-cycle will take the form of a total angle of rotation:

$$\int_{z_1} \omega = i \int_0^{2\pi} d\theta = 2\pi i \,. \qquad (3.10)$$

If $N$ were not encircled by $z_1$, in such a way that $z_1$ bounded a 2-cycle $c_2$ (such as a disc) then one could use Stokes's theorem ([1]) to show that:

$$\int_{z_1} \omega = \int_{\partial c_2} \omega = \int_{c_2} d_\wedge \omega = \int_{c_2} \Omega = 0. \qquad (3.11)$$

Hence, the non-vanishing of the holonomy integral in (3.10) is a symptom that perhaps the singular point $N$ is, in some sense, a possible source of curvature.

*c. Extending the connection to a singular point.* – As a first attempt at extending the connection $\omega$ to the singular point $N$, let us first define a step function at $N$ that takes the form of a function of $r$ that we shall call $[r]_N$ with the property that $[0]_N = 0$ and $[r]_N = 1$ when $r > 0$. Clearly, it is not continuous (much less differentiable), but if we were to treat it as a distribution then we would be justified in saying that it could be differentiated, and in fact:

$$d\,[r]_N = \delta(0)\, dr = -\,\delta(0)\, \theta^1, \qquad (3.12)$$

in which $\delta(0)$ is the Dirac delta distribution that is centered on $r = 0$.

We then define the extension of $\omega$ by way of:

$$\omega_N = [r]_N \, \omega = \begin{cases} 0 & \text{at } N \\ \omega & \text{otherwise.} \end{cases} \qquad (3.13)$$

---

([1]) As an historical aside, it is interesting that as one goes back in the literature of mathematics towards the era in which Stokes "proved' his theorem, an increasing number of authors insist that Stokes merely learned about it from Kelvin, who properly deserved the credit for first proving it.

In other words, we have simply defined the value of the connection 1-form to be zero at the singular point.

Upon exterior differentiation, one will get:

$$\Omega_N = d_\wedge \omega_N = d\,[r]_N \wedge \omega + [r]_N\, d_\wedge \omega = -\frac{i}{r}\delta(0)\,\theta^1 \wedge \theta^2. \tag{3.14}$$

If one puts this into the form:

$$\Omega_N = i\,\delta(0)\,dr \wedge d\theta \tag{3.15}$$

then one can see that this amounts to a 2-form with imaginary values that behaves like a Dirac delta function at $N$. Indeed, this time, if one includes $N$ in the region $c_2$ that is bounded by a circle of radius $r_0$, $z_1 = \partial c_2$, then the integral in (3.11) will become:

$$\int_{z_1} \omega_N = \int_{c_2} \Omega_N = i\int_0^{2\pi}\int_0^{r_0} \delta(0)\,dr \wedge d\theta = 2\pi i, \tag{3.16}$$

which equals the value (3.10) in the case where $N$ is missing from the region encircled by $z_1$.

Since analogous reasoning can be applied to the South pole, one can combine the two extensions into:

$$\bar{\omega} = [r]\,\omega, \tag{3.17}$$

where this time:

$$[r] = \begin{cases} 0 & r = 0,\,\pi R \\ 1 & \text{otherwise} \end{cases}, \tag{3.18}$$

in which $R$ represents the radius of the sphere. Differentiation gives:

$$d\,[r] = [\delta(0) + \delta(\pi R)]\,dr = -[\delta(0) + \delta(\pi R)]\,\theta^1, \tag{3.19}$$

which makes:

$$\bar{\Omega} = d_\wedge \bar{\omega} = \Omega_N + \Omega_S = -\frac{i}{r}[\delta(0) + \delta(\pi R)]\,\theta^1 \wedge \theta^2, \tag{3.20}$$

and if we write $\bar{\Omega}$ in the form:

$$\bar{\Omega} = i\,[\delta(0) + \delta(\pi R)]\,dr \wedge d\theta \tag{3.21}$$

then the integral of $\bar{\Omega}$ over all of $S^2$ will give:

$$\int_{S^2} \bar{\Omega} = \int_{S^2} \Omega_N + \int_{S^2} \Omega_S = 2\pi i + 2\pi i = 4\pi i. \tag{3.22}$$

Since the Euler-Poincaré characteristic of the 2-sphere is $\chi\,[S^2] = 2$, one has an example of the *Gauss-Bonnet theorem* in this calculation

$$\frac{1}{2\pi i} \int_{S^2} \bar{\Omega} = \chi [S^2] . \tag{3.23}$$

The expression on the left-hand side relates to the "first Chern class" of the $U(1)$-principle bundle that is defined by the bundle of oriented, orthonormal frames on $S^2$. We shall not elaborate here, but refer the interested reader to Milnor and Stasheff [7] again. Indeed, before Chern came up with his celebrated classes, he first looked into extending the Gauss-Bonnet theorem to even-dimensional manifolds of dimension greater than two.

*d. Parallel translation and geodesics.* – Suppose that $\gamma(s)$, $0 < s < 1$, is a smooth curve in $S^2$ and $\mathbf{v}(s)$ is a vector field along $\gamma$.

If the curve $\gamma$ does not pass through any singular points of $\mathbf{e}_i$ then one can define parallel translation of the vector $\mathbf{v}(s)$ along the curve $\gamma(s)$ with respect to the frame field $\mathbf{e}_i(s)$ by saying that $\mathbf{v}(s)$ must have the same components with respect to each frame along the curve. If $\mathbf{v}(s) = v^i(s)\,\mathbf{e}_i(s)$ then that condition amounts to:

$$v^i(s) = v_0^i \qquad \text{or} \qquad \frac{dv^i}{ds} = 0 \qquad \text{for all } s. \tag{3.24}$$

When the curve $\gamma(s)$ passes through a singular point, such as $N = \gamma(s_N)$, one cannot define the components of $\mathbf{v}(s)$ at the singular point, since a frame does not exist at it. However, as long as one is demanding the continuity of the vector field at the singular point, one can define $\mathbf{v}(s_N)$ to be limit that $\mathbf{v}(s)$ approaches as $s$ approaches $s_N$ :

$$\mathbf{v}(s_N) = \lim_{s \to s_N} \mathbf{v}(s) . \tag{3.25}$$

In particular, this also works for extending geodesics of the teleparallelism connection to singular points, since a geodesic of the teleparallelism connection is defined to be a curve whose velocity vector field is parallel along the curve; hence, its components must be constant with respect to the chosen frame field. Furthermore, since frame fields are composed of vector fields, one can similarly extend a frame field along a curve to a singular point by continuity. However, since and infinitude of curves will converge to s singular point, that does not imply that either a vector field or a frame field can be extended to the singular point in a consistent way.

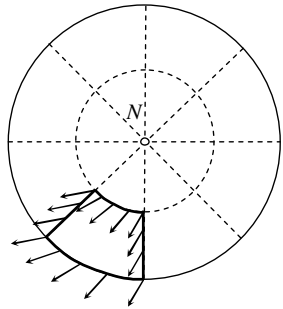 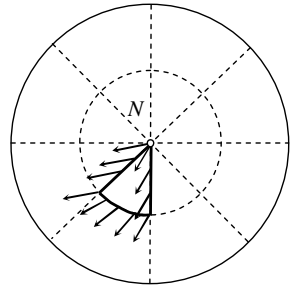

Figure 4.*a*. – Parallel translation around a loop that does not go through *N*.

Figure 4.*b* – Parallel translation around a loop that does go through *N*.

One can see the fundamental difference between loops that pass through a singular point (such as *N*) and ones that do not in the figures above. In Fig. 4.*a*, one has parallel translation around a loop that does not go through *N*, and one can see that the vector will return to itself at the end of the loop. By contrast, in Fig. 4.*b*, when the loop passes through *N*, the effect of parallel translation around the loop will be to return the vector with a rotation through the angle between the two longitude circles that intersect at *N*. Hence, in the eyes of holonomy, the singular point at *N* has introduced curvature.

One sees that the actual rotation of the vector happens along the latitude circles, so since the two latitude arcs in Fig. 4.*a* involve the same change in longitude, not net rotation of the vector will occur under parallel translation. However, in Fig. 4.*b*, since only one latitude circle is involved with the loop, the net effect on the vector when it returns to *N* will be to rotate it through an angle that equals the change in longitude between the two longitude circles upon which the other two arcs of the spherical triangle lie.

*e. Smoothing out the discontinuity.* – Now, let us try to smooth out the discontinuity that we have introduced into the connection and its curvature. The way that we shall do that is by way of a smooth "bump" function ([1]), which is a smooth function $\rho(r)$ such that $\rho(0) = 0$, $\rho(r) = 1$ when $r > r_0$, which is some sufficiently small number that the circle about *N* (or *S*) of radius $r_0$ does not include any other singular points, and $\rho(r)$ is smooth and monotonically-increasing between $r = 0$ and $r = r_0$. By contrast, its derivative $d\rho/dr$ is equal to 0 when $r = 0$ and $r > r_0$ and positive in between those limits. We illustrate those functions in Fig. 5:

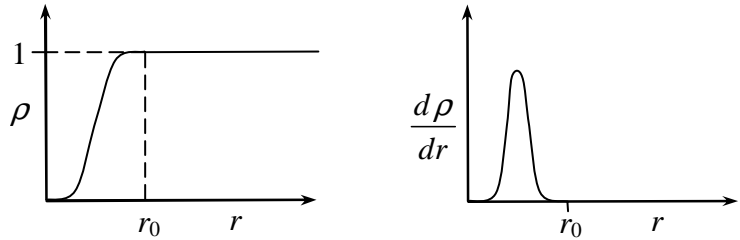

Figure 5. – A smooth bump function and its derivative.

One can see that as $r_0$ approaches 0, the bump function will approach a step function centered on 0, while its derivative will approach a delta function centered on 0.

If we now define the bump function centered on *N* to be $\rho_N(r)$ then the extension of $\omega$ this time will be:

$$\omega_N = \rho_N \, \omega . \qquad (3.26)$$

Exterior differentiation gives:

$$\Omega_N = d_\wedge \omega_N = d\rho_N \wedge \omega = i \, d\rho \wedge d\theta , \qquad (3.27)$$

---

([1]) Also called an "Urysohn function." Such things will always exist as long as *M* is regular, but typically one assumes that differentiable manifolds are normal, which is a stronger separation axiom. (One might confer Munkres [**9**] for basic topological definitions.) The reader can find an explicit example of a smooth bump function in Guillemin and Pollack [**10**].

and when this is integrated over a 2-chain $c_2$ that includes $N$, one will get:

$$\int_{c_2} \Omega_N = i \int_0^{2\pi}\int_0^1 d\rho \wedge d\theta = 2\pi i, \qquad (3.28)$$

which is the same value that we got from the delta function formulation.

The extension to a curvature 2-form $\bar{\Omega}$ that smoothly approximates the one in (3.20) is clear from here, and one will once again get the Gauss-Bonnet theorem upon integrating it over the entire sphere.

From this example, we can say that in the case of the 2-sphere, extending the teleparallelism connection, whose curvature must vanish, to a connection on a non-parallelizable manifold is certainly straightforward in the case of a finite set of isolated singular points. Furthermore, the singular points can also be interpreted as the sources of non-vanishing curvature.

**4. Extending the teleparallelism connection for more general singular frame fields.** – When one goes over the preceding discussion, one sees that only some of the statements were specific to the 2-sphere, while the basic construction could be easily extended to higher dimensions. Hence, we shall now extend the techniques to $n$-dimensional differentiable manifolds that admit singular frame fields with a finite number of isolated points for their singularities.

*a. The basic construction.* – Suppose that $\{\mathbf{e}_i, i = 1, …, n\}$ is a generic singular frame field on an $n$-dimensional manifold, and let $Z(\mathbf{e})$ be its singularity set, which is assumed to consist of a set of isolated points; let $N$ be one of those points.

The complement $Z^c(\mathbf{e})$ of $Z(\mathbf{e})$ in $M$ is a parallelizable open subset of $M$ (and therefore a differentiable manifold), so the restriction of $\mathbf{e}_i$ to $Z^c(\mathbf{e})$ will define a global frame field on $Z^c(\mathbf{e})$, and with it, a teleparallelism connection $\omega^i_j$, along with its associated torsion $\Theta^i$ and (vanishing) curvature $\Omega^i_j$. (From now on, we occasionally shall skip the explicit reference to the indices, for the sake of clarity.)

Choose an open neighborhood $B(r_0)$ of $N$ that is diffeomorphic to an open $n$-ball of radius $r_0$ and is sufficiently small that it does not contain any other singular points. Define a smooth bump function $\rho(r)$ on $B(r_0)$ that approaches 1 as $r$ approaches $r_0$ and 0 as $r$ approaches 0, while being monotonically increasing from 0 to 1 as $r$ increases. Hence, one is still dealing with the qualitative picture that is depicted in Fig. 5.

One can then multiply the connection 1-form $\omega$ (suitably extended with zero at $N$) by the bump function $\rho$ and obtain a form $\bar{\omega}$ that extends $\omega$ to the point $N$:

$$\bar{\omega} = \rho\, \omega. \qquad (4.1)$$

One immediately sees that when $\rho = 1$, $\bar{\omega} = \omega$ and when $\rho = 0$, $\bar{\omega} = 0$.

*b. The structure equations of the extended connection.* – In order to evaluate the structure equations for the connection $\bar\omega$, one must first decide upon what to do about extending the frame field **e**$_i$ to *N*. That is where it is more convenient to regard the singular points as points where some of the frame members vanish, rather than points where they are not defined, since in the former case, no extension of the field is necessary. Hence, we shall assume that the frame field **e**$_i$ is singular at *N* in the sense of vanishing frame members.

With the substitution (4.1), the structure equation for the torsion of $\bar\omega$:

$$\bar\Theta = d_\wedge \theta + \bar\omega \wedge \theta \tag{4.2}$$

will then take the form:

$$\bar\Theta = \rho\,\Theta + d\rho \wedge \theta. \tag{4.3}$$

Meanwhile, the structure equation for curvature:

$$\bar\Omega = d_\wedge \bar\omega + \bar\omega \wedge \bar\omega \tag{4.4}$$

will become:

$$\bar\Omega = d\rho \wedge \omega - \rho(1-\rho)\,\omega \wedge \omega. \tag{4.5}$$

The fact that $\Omega$ must vanish on its domain of definition has been used in this.

One sees that since $\rho = 0$ or 1 and $d\rho = 0$ at the center and surface of the ball $B(r_0)$, one will have

$$\bar\Theta = \begin{cases} \Theta & \text{for } \rho = 1 \\ 0 & \text{for } \rho = 0 \end{cases}, \qquad \bar\Omega = 0 \ (\rho = 0 \text{ or } 1). \tag{4.6}$$

Hence, the torsion and curvature of the extended connection 1-form $\bar\omega$ do, in fact, agree with their pre-existing definitions individually.

Of particular interest is the fact that $\bar\Omega$ will not generally vanish in the interior of the ball $B(r_0)$, since its vanishing would give:

$$d\rho \wedge \omega - \rho(1-\rho)\,\omega \wedge \omega = 0. \tag{4.7}$$

Since this could not be true for all possible choices of $\rho$ identically, one must infer that there will be some choices of $\rho$ for which $\bar\Omega$ is non-vanishing.

Therefore, one is justified in saying that the singular point at *N* has become the source of non-zero curvature in the extended connection. (Of course, the curvature will still be zero at *N*.)

As long as one is dealing with a singularity set $Z(\mathbf{e})$ that consists of isolated points, one can say that the same constructions and considerations can be applied to each of them. Hence, one can obtain an extended connection $\bar\omega$ that is defined on all of *M*.

*c. Parallel translation and geodesics of the extended connection.* – Parallel translation under the extended connection still works the same way as in the example of $S^2$. That is, when a smooth curve $\gamma(s)$ in *M* does not pass through any singular points, a vector field **v** (*s*) along that curve will be parallel for the teleparallelism connection iff its components $v^i$ (*s*) with respect to the frame field $\mathbf{e}_i$ (*s*) along that curve are constants. When the curve does pass through a singular point $N = \gamma(s_N)$, one defines the extension of **v** (*s*) to *N* by continuity, as before, and similarly for frame fields along $\gamma$.

Once again, this allows one to extend the definition of the geodesics of the teleparallelism connection to the extended connection that includes the singular points.

The definition of parallel translation of higher-rank tensor fields along curves is simply the statement that their components with respect to the basis on their vector space that is defined by all tensor products $\mathbf{e}_{i_1} \otimes \cdots \otimes \mathbf{e}_{i_p} \otimes \theta^{j_1} \otimes \cdots \otimes \theta^{j_q}$ must be constant.

*d. Extending the volume element and metric.* – The same basic construction that was used to extend the teleparallelism connection to the singular points can also be applied to the volume element and metrics that are defined by a frame field, except that the results will typically be more singular than is necessary.

Recall (from [**8**], say) that the volume elements that are defined by a frame field $\mathbf{e}_i$ and its reciprocal coframe field $\theta^i$ are the *n*-vector field **V** and the *n*-form *V* that one gets from:

$$\mathbf{V} = \mathbf{e}_1 \wedge \ldots \wedge \mathbf{e}_n = \frac{1}{n!} \varepsilon^{i_1 \cdots i_n} \mathbf{e}_{i_1} \wedge \cdots \wedge \mathbf{e}_{i_n}, \tag{4.8}$$

$$V = \theta^1 \wedge \ldots \wedge \theta^n = \frac{1}{n!} \varepsilon_{i_1 \cdots i_n} \theta^{i_1} \wedge \cdots \wedge \theta^{i_n}, \tag{4.9}$$

respectively.

It is easy to see by inspection that these tensor fields are parallel under the connection that makes the frame field parallel, since they both have constant components (namely, 1) with respect to the basis that is defined by $\mathbf{e}_1 \wedge \ldots \wedge \mathbf{e}_n$, in one case, and $\theta^1 \wedge \ldots \wedge \theta^n$, in the other.

If one extends the volume element by way of:

$$\overline{V} = \rho V \tag{4.10}$$

then one will find that since $\overline{V} = 0$ at the singular point about which the bump function $\rho$ is centered, it will not be a true volume element, anymore. On the other hand, if *M* (or really, its tangent bundle) is orientable then a global non-zero *n*-form should exist on it, even when *M* is not parallelizable. Hence, in a sense, the *n*-form $\overline{V}$ is not always indicative of the topology of the manifold in the same way that the extended connection is.

One can define a metric *g* of any signature type $\eta_{ij}$ = diag [− 1, …, − 1, + 1, …, +1] on a parallelizable manifold with a given global frame field $\mathbf{e}_i$ by demanding that it should make the global frame field orthonormal for that signature type. One then gets:

$$g = \eta_{ij} \theta^i \theta^j, \tag{4.11}$$

in which the unwritten product is a symmetrized tensor product.

Once again, if one makes the obvious extension to a singular point:

$$\bar{g} = \rho \, g \tag{4.12}$$

then one will find that since $g = 0$ at the singular point, it will no longer be non-degenerate at that point, and will not define a metric on the tangent vectors to the singular point. Since any paracompact differentiable manifold admits a Riemannian metric (of signature type $\delta_{ij}$), one again has to accept that the extended metric $\bar{g}$ does not reflect the true metrizability of the manifold.

In both cases – viz., volume elements and metric – there are other techniques for extending partial volume elements and metrics to global ones, but since the main objective of the present study is to show that a certain technique for extending the teleparallelism connection to singular points can produce non-vanishing curvature, we shall not discuss the other techniques.

**5. Suspensions of parallelizable manifolds.** – The 2-sphere is an example of a more general construction, namely, a *suspension* of a manifold *M* (which shall be assumed to have no boundary). The basic construction that gives one the suspension of *M*, which shall be denoted by *SM*, is to first form the product manifold $M \times [0, 1]$ and then identify each boundary component $M \times \{0\}$ and $M \times \{1\}$ to a (separate) point, which we shall call *S* and *N*, respectively. Call the set of equivalence classes under that identification *SM*. Hence, there is a canonical projection $M \times [0, 1] \to SM$ that takes (*x*, *s*) to *S* when $s = 0$, *x* when $0 < s < 1$, and *N* when $s = 1$. One then gives *SM* the quotient topology, which is the finest (i.e., most open subsets) topology that will make the projection continuous, and a differential structure that mostly involves adding charts about *S* and *N*.

Any *n*+1-sphere is the suspension of the *n*-sphere: $S^{n+1} = SS^n$. In particular the 2-sphere is the suspension of a circle, $S^4$ is the suspension of $S^3$, and $S^8$ is the suspension of $S^7$. In each case, one is suspending a parallelizable manifold and obtaining a non-parallelizable one, and in each case, the singularity subset for the resulting singular frame field consists of two points, namely, *S* and *N*. That is because if *M* is parallelizable then the intermediate manifold $M \times [0, 1]$ will also be so, and the singularities in the frame field will arise only by identifying all of the points of $M \times \{0\}$ to *S* and all of the points of $M \times \{1\}$ to *N*.

One naturally wonders how suspension affects the parallelizability of a manifold, in general. In fact, one can start with a fairly general:

**Theorem:**

*If a topological space X has an Euler-Poincaré characteristic that equals $\chi[X]$ then its suspension will have an Euler-Poincaré characteristic that equals $1 - \chi[X]$.*

**Proof:**

One starts with the general theorem from singular homology (see, [**11**], in particular) that if the integer singular homology modules of $X$ are $H_k(X)$, $k = 0, 1, \ldots, n$ then the homology modules of its suspension $SX$ will be:

$$H_0(SX) = \mathbb{Z}, \quad H_k(SX) \cong H_{k-1}(X), k = 1, \ldots, n+1.$$

Since the $k^{\text{th}}$ Betti number $b_k$ of $X$ is the rank of the free part of $H_k(X)$, the $(k+1)^{\text{th}}$ Betti number of $H_k(SX)$ will also be $b_k$, while the $0^{\text{th}}$ Betti number of $SX$ will be 1. The Euler-Poincaré characteristic of $SX$ will then:

$$\chi[SX] = 1 - \sum_{k=0}^{n}(-1)^k b_k = 1 - \chi[X].$$

This has three useful corollaries:

**Corollary:**

*If $\chi[X]$ vanishes then $\chi[SX]$ cannot vanish. In particular, $\chi[SX] = 1$.*

**Corollary:**

*If a differentiable manifold admits a non-zero vector field then its suspension will not admit one.*

**Corollary:**

*If a differentiable manifold is parallelizable then its suspension will not be parallelizable.*

Since any compact, orientable 3-manifold is parallelizable, its suspension can serve as an interesting class of compact, orientable, non-parallelizable 4-manifolds. A four-torus is one example of a compact, orientable, parallelizable 4-manifold, so not all compact, orientable 4-manifolds are non-parallelizable.

Basically, if $\{\mathbf{e}_1, \ldots, \mathbf{e}_n\}$ is a frame field on $M$ and $\mathbf{e}_{n+1}$ is a non-zero vector field on $[0, 1]$ then the set $\{\mathbf{e}_1, \ldots, \mathbf{e}_n, \mathbf{e}_{n+1}\}$ will be a frame field on $M \times [0, 1]$. In order to define the frames on the boundary components, one first thinks of $[0, 1]$ as being contained in the open interval $(a, b)$, so if $M$ has no boundary then neither will $M \times (a, b)$. The vector field $\mathbf{e}_{n+1}$ can then be extended differentiably to $(a, b)$ in some way. One can then think of the tangent spaces to the boundary points as being $n+1$-dimensional, in such a way that the tangent spaces to the boundary itself will be $n$-dimensional subspaces of those $n+1$-dimensional spaces. Since differentiation is a purely local process, it will not matter what choice of $(a, b)$ or what extension of $\mathbf{e}_{n+1}$ that one makes, because the restriction of the extension to $M \times [0, 1]$ will be the same in any case.

It is clear that the frame fields $\mathbf{e}_i(x, 0)$ on $M \times \{0\}$ and $\mathbf{e}_i(x, 1)$ on $M \times \{1\}$ cannot generally be identified with a single frame in either case. Hence, although the canonical projection $M \times [0, 1] \to SM$ will take each point between the boundary components to itself diffeomorphically, so the complement of $\{N, S\}$ in $SM$ will be diffeomorphic to $M \times (0, 1)$ and one can map the frames on $M \times (0, 1)$ to corresponding frames on $SM - \{N, S\}$, the same thing cannot be said of the frames on the boundary components. Hence, one must treat the points $N$ and $S$ as singular points of the frame field on $SM$ that is defined by projection, in the sense that the resulting frame field is not defined at those points. One can also allow the vector fields to go to zero on the boundary components, possibly with the use of bump functions, which is more consistent with our general procedure for extending the teleparallelism connection.

This process is illustrated in the following figure:

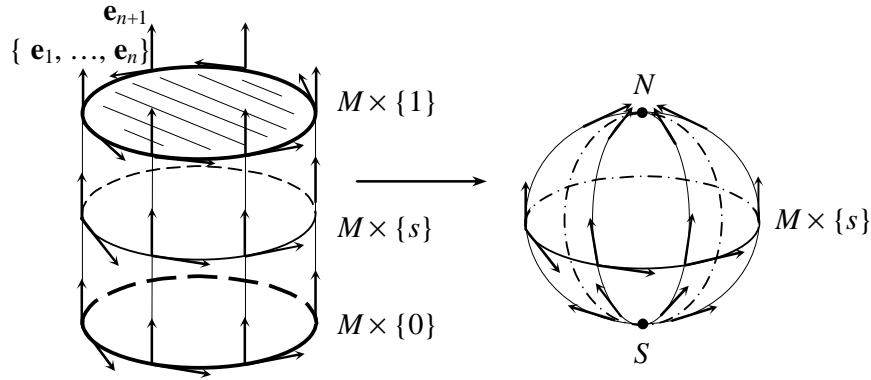

Figure 6. – Defining a singular frame field by suspension.

Thus, what we have defined by projection onto $SM$ is a singular frame field whose singularity set consists of two isolated points. Hence, the methods of Section **4** can be applied, and one will then obtain a connection on $SM$ that is defined by the singular frame field that is obtained by projection, and the curvature of that connection will generally be non-vanishing in the neighborhood of the singular points.

In particular, this construction can be applied to the 2-sphere and the 4-sphere, both of which are used in physical field theories when one wishes to deal with fields on the plane or $\mathbb{R}^4$ that "vanish at infinity." The basic construction in such a case is usually stereographic projection of the sphere onto the plane or $\mathbb{R}^4$, as the case may be.

**6. Discussion.** – As mentioned above, although the technique of bump functions is sufficient to show the existence of an extended connection and the possibility that its curvature might vanish, nonetheless, there is something less than canonical about it. Hence, one direction for further development of the theory would be towards finding those cases in which such a singular teleparallelism connection might exist on a non-parallelizable manifold in a more canonical way.

Another direction of growth for the technique is to extensions of the teleparallelism connection when the singularity set of a generic singular frame field consists of more than just isolated points. Since a set of isolated points is an elementary example of a 0-dimensional simplicial complex, the obvious generalization would be the one that Stiefel made, namely, an *n*-dimensional simplicial complex.

Since no mention has been made in this article of the application of this singular teleparallelism connection to physical field theories, that would clearly be an important direction to pursue. Although the discovery of gravito-electromagnetism has certainly cast a whole new light on the validity of the original Einstein-Maxwell unification problem by showing the Maxwell equations to be typical of weak-field theories, nonetheless, extensions of Maxwell's equations to strong field theories exist in various forms. Of particular interest is the formulation of Einstein's equations for gravitation in the language of teleparallelism [**12**, **13**].

Furthermore, as the author has pointed out before [**14**], the geometry of teleparallelism is essentially an extension of the Frenet-Serret equations of moving frames along curves, which are directly related to the kinematics of bending and twisting of one-dimensional objects in continuum mechanics, to the bending and twisting of higher-dimensional objects, so the extension to singular teleparallelism might be relevant to that study, as well. In particular, one might consider the role of topological defects, such as dislocations and disclinations, in generating curvature.

____________